\documentclass[11pt,twoside]{article}

\usepackage{asp2014}

\aspSuppressVolSlug
\resetcounters

\begin{document}

\title{The Challenges of a Public Data Release: behind the scenes of SDSS DR13}

\author{Anne-Marie Weijmans,$^1$ Michael Blanton,$^2$ Adam S. Bolton,$^{3,4}$
  Joel Brownstein,$^4$ M. Jordan Raddick,$^5$ and Ani Thakar$^5$
\affil{$^1$University of St Andrews, St Andrews, Scotland, UK; \email{amw23@st-andrews.ac.uk}}
\affil{$^2$New York University, New York, New York,
  US}
\affil{$^3$NOAO, Tucson,  Arizona, US}
\affil{$^4$University of Utah, Salt Lake City, Utah,
  US}
\affil{$^5$The Johns Hopkins University, Baltimore, Maryland,
  US}
}

\paperauthor{Anne-Marie
  Weijmans}{amw23@st-andrews.ac.uk}{}{University of St
  Andrews}{School of Physics and Astronomy}{St Andrews}{Fife}{KY16 9SS}{UK}
\paperauthor{Michael Blanton}{Blanton@physics.nyu.edu}{}{New York
  University}{Department of Physics}{New York}{New York}{NY 10003}{US}
\paperauthor{Adam S. Bolton}{bolton@noao.edu}{}{NOAO}{Community Science and Data Center}{Tucson}{Arizona}{AZ 85719}{US}
\paperauthor{Joel Brownstein}{joelbrownstein@astro.utah.edu}{}{University of
  Utah}{Department of Physics and Astronomy}{Salt Lake City}{Utah}{UT 84112}{US}
\paperauthor{M. Jordan Raddick}{raddick@jhu.edu}{}{Johns Hopkins University}{Institute for Data Intensive Engineering and Science}{Baltimore}{Maryland}{MD 21218}{US}
\paperauthor{Ani Thakar}{thakar@jhu.edu}{}{Johns Hopkins University}{Institute for Data Intensive Engineering and Science}{Baltimore}{Maryland}{MD 21218}{US}

\begin{abstract}
  The Sloan Digitial Sky Surveys (SDSS) have been collecting imaging
  and spectoscopic data since 1998. These data as well as their
  derived data products are made publicly available through regular
  data releases, of which the 13th took place summer 2016. Although
  public data releases can be challenging to manage, they signficantly
  increase the impact of a survey, both scientifically and educationally.
\end{abstract}

\section{Introduction}
The Sloan Digital Sky Survey (SDSS) started observing in 1998, with
the goal to image the Northern hemisphere in five optical wavelength
bands \citep{2000AJ....120.1579Y}, using the dedicated 2.5m Sloan
Foundation Telescope \citep{2006AJ....131.2332G} at Apache Point
Observatory (APO), New Mexico, US. Since then, SDSS has hosted several
dedicated surveys, adding optical, infrared and integral-field
spectroscopy to its orignal imaging survey
\citep[e.g.][]{2011AJ....142...72E}. The current SDSS-IV (Blanton et
al. in prep) consists of three spectroscopic surveys: i) APOGEE-2
(Apache Point Observatory Galaxy Evolution Experiment), obtaining
infrared spectra of stars to unravel the chemical and dynamical
formation history of the Milky Way (Majewski et al. in prep), ii)
MaNGA (Mapping Nearby Galaxies at APO), an integral-field
spectroscopic survey to study the formation, growth and evolution of
galaxies \citep{2015ApJ...798....7B}, and iii) eBOSS (extended Baryon
Oscillation Spectroscopic Survey), measuring redshifts of $\sim$1.5
million galaxies and quasars from optical spectra to map the
structure of the Universe and determine its expansion history
\citep{2016AJ....151...44D}. In 2017, SDSS will for the first time
observe from the Southern hemisphere at Las Campanas Observatory
(LCO), Chile, as part of APOGEE-2. The SDSS collaboration includes
more than 50 member institutes spread over 4 continents, with close
to 1000 scientists registered as members.

SDSS has from its beginnings been dedicated to making its data
publicly available through data releases. The first early data release
was made public in 2002 \citep{2002AJ....123..485S}, and in summer 2016
the Thirteenth SDSS data release (DR13) became publicly available
\citep{2016arXiv160802013S}. These public data releases have
signficantly widened the impact of the SDSS beyond its
collaborations: more than 7000 papers have been published based on
SDSS data. The 2015 National
Research Council report ``Optimizing the U.S. Ground-Based Optical and
Infrared Astronomy System'' notes that within this US system the
SDSS has a publication rate three times higher than any other
telescope. In addition, SDSS data has been incorporated in various
educational (e.g., Voyages\footnote{http://voyages.sdss.org}) and
public engagement activities \citep[e.g., Galaxy
Zoo;][]{2008MNRAS.389.1179L}. In the field of data science, SDSS
public data releases have been cited as the most influential data
source ahead of any other source in any field of science \citep{stalzer16}. In
this proceeding we discuss the mechanisms of an SDSS public
data release, with a focus on the latest release -  SDSS-IV's DR13.

\section{Data and Data Products}

The main components of a data release are the data products.
The raw SDSS data is transferred daily from APO to the Center for High
Performance Computing (CHPC) at the University of Utah, where it is
reduced by dedicated data reduction pipelines. Each survey within
SDSS-IV has a survey data team that is responsible for this part of
the process. Once the data is reduced and the relevant data products
have been vetted by the survey data teams, they are copied to the
Science Archive Server (SAS)\footnote{http://data.sdss.org/sas} where they remain as proprietary data for
the SDSS-IV collaboration, until they are incorporated in a public
data release (average time scale $\sim$1.5 year).

The SAS offers a file-based system that allows for low-level bulk data
access and off line analysis. The SAS can also be accessed through a
specialised web application, to visualise spectra. In addition, the
Catalog Archive Server (CAS)\footnote{http://skyserver.sdss.org}, hosted at the Johns Hopkins University,
offers Web browser-based graphical and SQL query interfaces to the
catalog data that permit casual synchronous data retrieval
\citep[SkyServer;][]{Szalay:2002:SSP:564691.564758}
as well as batch-mode server-side analysis \citep[CasJobs;][]{2008CSE....10...18L}
with advanced capabilities like a server-side personal SQL database,
data upload and data sharing via groups. There is also a command-line
tool to submit CasJobs queries, and as of DR13, a jupyter notebook
interface to retrieve catalog data from SkyServer and CasJobs
within a docker container. Both
the SAS and the CAS offer public data access, with the SAS hosting 267
TB of data products, as of DR13. All data releases are cumulative, in
that they also contain data products from previous data releases. By
accessing the latest SDSS data release, users will therefore always
have access to all available SDSS public data.

\section{Documentation}

Making data publicly available is a necessary but not sufficient
condition for generating impact: in addition, the data has to be
\emph{accessible}. Users need to know where to find the data, and how
to work with the data. Documentation is essential, and each SDSS data
product therefore has a detailed data model. Each data release is also
described in detail in an accompaning data release paper \citep[for DR13,
see][]{2016arXiv160802013S}. 

However, a public data release serves many different
audiences. Although for astronomers familiar with the SDSS data format
the data model may be sufficient, novel users and students require
more documentation. The SDSS website\footnote{http://www.sdss.org/}
therefore offers a portal with more information on how to access the
data (including links to the SAS, CAS and web application), as well as
background information, links to technical papers, examples and
tutorials. As teachers and instructors also use SDSS data in their
classrooms, the website also offers links to education and public
engagement activities. Finally, users can e-mail the SDSS helpdesk if
they have any questions or encounter any problems when working with
SDSS data.

\section{Data Release Management}

The SDSS-IV data team is led by the Science Archive Scientist, the
Catalog Archive Scientist and the Data Release Coordinator. The latter
is repsonsible for managing the data release and keeping it on
schedule. Planning for a public data release typically starts one year
in advance. First the data products are generated by the survey data
teams, and vetted for quality control by the survey science teams. Once the data products
are finalised, catalogs are generated and loaded into the CAS. 

Most of the documenation is written during a one-week documentation
workshop (''DocuFeest''), $\sim$four months before the data release. Several members of the data team, the
survey teams and the education and public engagement team meet to
generate most of the website content, as well as a comprehensive draft
of the data release paper. Descriptions and tutorials for new data
products have to be added into the documenation hierarchy: for DR13
this meant a redesign of the website to include the new MaNGA
survey.  The website and paper are finalised in the weeks after the
DocuFeest. Meeting in person has proved crucial for efficient writing
of documentation, as both the website and paper require input from all
teams across SDSS-IV, and communication is key.

As the date for the data release approaches, a plan is drafted for the
announcement of the data release. SDSS-IV has a strong social media
presence, with a general and survey-specific Twitter accounts, a
Facebook page, and a blog aimed at the general public. For DR13, we
issued a Facebook announcement and tweeted in almost all languages spoken within the
collaboration (thanks to a team-wide effort to translate the tweets). A description of the content of
DR13 was given in a multilingual blog post. Announcements were
also issued over mailing lists aimed at professional astronomers.

\section{Conclusion}
The impact of an astronomical survey is set by the reach of its data
distribution system. If the data does not reach the astronomers for
their research projects to make new discoveries, the teachers to teach
their students how to work with astronomical data, and the general
public to increase their awareness of astronomy and science, then
impact will be limited to a small core survey team. A successful
public data release, especially one aimed at a variety of end users,
therefore needs to ensure that the data is not only freely available,
but also clearly documented and in accessible formats. 

Managing a public data release is challenging: it involves keeping track of a
myriad of different data products and their documenation, which
includes data models, descriptions and tutorials, all aimed at
different user audiences. Quality control of both the data products
and the documentation is crucial. Good communication between different
teams within the collaboration is key, given that the data release
requires input from everyone. But despite the challenges and work
involved, public data releases are worthwhile: they increase the
impact of the survey, allow for more science output beyond the
collaboration, and provide accessible data products not only for
astronomical research, but also for education and public engagement.

\acknowledgements 
AW acknowledges support of a Leverhulme Trust Early Career Fellowship. The
authors would like to thank the  Center for Data Science at New York University for their
hospitality during DocuFeest 2016. Funding for the Sloan Digital Sky Survey IV has been provided by
the Alfred P. Sloan Foundation, the U.S. Department of Energy Office of
Science, and the Participating Institutions. SDSS-IV acknowledges
support and resources from the Center for High-Performance Computing at
the University of Utah. The SDSS web site is www.sdss.org. SDSS-IV is managed by the Astrophysical Research Consortium for the 
Participating Institutions of the SDSS Collaboration including the 
Brazilian Participation Group, the Carnegie Institution for Science, 
Carnegie Mellon University, the Chilean Participation Group, the French Participation Group, Harvard-Smithsonian Center for Astrophysics, 
Instituto de Astrof\'isica de Canarias, The Johns Hopkins University, 
Kavli Institute for the Physics and Mathematics of the Universe (IPMU) / 
University of Tokyo, Lawrence Berkeley National Laboratory, 
Leibniz Institut f\"ur Astrophysik Potsdam (AIP),  
Max-Planck-Institut f\"ur Astronomie (MPIA Heidelberg), 
Max-Planck-Institut f\"ur Astrophysik (MPA Garching), 
Max-Planck-Institut f\"ur Extraterrestrische Physik (MPE), 
National Astronomical Observatory of China, New Mexico State University, 
New York University, University of Notre Dame, 
Observat\'ario Nacional / MCTI, The Ohio State University, 
Pennsylvania State University, Shanghai Astronomical Observatory, 
United Kingdom Participation Group,
Universidad Nacional Aut\'onoma de M\'exico, University of Arizona, 
University of Colorado Boulder, University of Oxford, University of Portsmouth, 
University of Utah, University of Virginia, University of Washington, University of Wisconsin, 
Vanderbilt University, and Yale University.

\bibliography{O7-1}  

\begin{thebibliography}{}
\expandafter\ifx\csname natexlab\endcsname\relax\def\natexlab#1{#1}\fi
\expandafter\ifx\csname url\endcsname\relax
  \def\url#1{\texttt{#1}}\fi
\expandafter\ifx\csname urlprefix\endcsname\relax\def\urlprefix{URL }\fi
\providecommand{\eprint}[2][]{\url{#2}}

\bibitem[{{Albareti} et~al.(2016)}]{2016arXiv160802013S}
{Albareti}, F.~D., et~al. 2016, ArXiv e-prints. \eprint{1608.02013}

\bibitem[{{Bundy} et~al.(2015)}]{2015ApJ...798....7B}
{Bundy}, K., et~al. 2015, \apj, 798, 7

\bibitem[{{Dawson} et~al.(2016)}]{2016AJ....151...44D}
{Dawson}, K.~S., et~al. 2016, \aj, 151, 44

\bibitem[{{Eisenstein} et~al.(2011)}]{2011AJ....142...72E}
{Eisenstein}, D.~J., et~al. 2011, \aj, 142, 72

\bibitem[{{Gunn} et~al.(2006)}]{2006AJ....131.2332G}
{Gunn}, J.~E., et~al. 2006, \aj, 131, 2332

\bibitem[{{Li} \& {Thakar}(2008)}]{2008CSE....10...18L}
{Li}, N., \& {Thakar}, A.~R. 2008, Computing in Science and Engineering, 10, 18

\bibitem[{{Lintott} et~al.(2008)}]{2008MNRAS.389.1179L}
{Lintott}, C.~J., et~al. 2008, \mnras, 389, 1179

\bibitem[{{Stalzer} \& {Mentzel}(2016)}]{stalzer16}
{Stalzer}, M., \& {Mentzel}, C. 2016, SpringerPlus, 5, 1266

\bibitem[{{Stoughton} et~al.(2002)}]{2002AJ....123..485S}
{Stoughton}, C., et~al. 2002, \aj, 123, 485

\bibitem[{{Szalay} et~al.(2002)}]{Szalay:2002:SSP:564691.564758}
{Szalay}, A.~S., et~al. 2002, in Proceedings of the 2002 ACM SIGMOD
  International Conference on Management of Data (New York, NY, USA: ACM),
  SIGMOD '02, 570

\bibitem[{{York} et~al.(2000)}]{2000AJ....120.1579Y}
{York}, D.~G., et~al. 2000, \aj, 120, 1579

\end{thebibliography}

\end{document}